%% file: icstcc_v7.tex
\newcommand{\CLASSINPUTtoptextmargin}{56pt}
\newcommand{\CLASSINPUTbottomtextmargin}{72pt}
\newcommand{\CLASSINPUTinnersidemargin}{45pt}
\newcommand{\CLASSINPUToutersidemargin}{45pt}

\documentclass[conference,a4paper,final]{IEEEtran}
\IEEEoverridecommandlockouts
\usepackage{cite}
\usepackage{amsmath,amssymb,amsfonts}
\usepackage{algorithmic}
\usepackage{graphicx}
\usepackage{textcomp}
\usepackage{xcolor}
\usepackage{scrextend}
\usepackage[utf8]{inputenc}
\usepackage[T1]{fontenc}
\usepackage{changepage}
    
\usepackage{tikz}
\usepackage{mathptmx} % To emulate your `font`
\DeclareMathAlphabet{\mathcal}{OMS}{cmsy}{m}{n}
\usetikzlibrary{decorations.pathmorphing} % To decorate the surface of water
\usetikzlibrary{arrows.meta} % The newer options for arrows (PGF 3.0)
\usepackage[
			width=.8\linewidth,
      		]{caption} % 08_03_2014
\DeclareMathOperator*{\argmin}{arg\,min}

\definecolor{hellgelb}{rgb}{1,1,0.85}     % Hintergrundfarbe
\definecolor{colKeys}{RGB}{0,0,255}       % blau
\definecolor{colIdentifier}{RGB}{0,0,0}	  % schwarz
\definecolor{colComments}{RGB}{34,139,34} % gruen
\definecolor{colString}{RGB}{160,32,240}  % violett

\def\BibTeX{{\rm B\kern-.05em{\sc i\kern-.025em b}\kern-.08em
    T\kern-.1667em\lower.7ex\hbox{E}\kern-.125emX}}

\begin{document}

\changepage{27pt}{}{}{}{}{-27pt}{}{}{}

\title{System identification of a hysteresis-controlled pump system using SINDy
\thanks{* This research was funded by the German Ministry for Economy and Energy in context of the project \textit{EnEffReg}, 03ET1313B and partially extended in the project \textit{Reinforcement Learning for complex automation technology}, supported by Investitionsbank Berlin (IBB), co-financed by the European Regional Development Fund.}
}

\author{
\IEEEauthorblockN{Gregor Thiele and Arne Fey}
\IEEEauthorblockA{Fraunhofer Institute for Production Systems\\ and Design Technology IPK\\
Department for process automation\\ and robotics,
Berlin, Germany \\
ORCID: 0000-0002-7108-5203}
\and
\IEEEauthorblockN{David Sommer}
\IEEEauthorblockA{Weierstrass Institute\\
for Applied Analysis and Stochastics\\
Berlin, Germany \\
ORCID: 0000-0002-6797-8009}
\and

\IEEEauthorblockN{Jörg Krüger}
\IEEEauthorblockA{
Institute for Machine Tools\\ and Factory Management\\
Technische Universität Berlin\\
Berlin, Germany\\
ORCID: 0000-0001-5138-0793
}
}

\maketitle

\begin{abstract}
Hysteresis-controlled devices are widely used in industrial applications. For example, cooling devices usually contain a two-point controller, resulting in a nonlinear hybrid system with two discrete states. Dynamic models of systems are essential for optimizing such industrial supply technology. However, conventional system identification approaches can hardly handle hysteresis-controlled devices. Thus, the new identification method Sparse Identification of Nonlinear Dynamics (SINDy) is extended to consider hybrid systems. SINDy composes models from basis functions out of a customized library in a data-driven manner. For modeling systems that behave dependent on their own past as in the case of natural hysteresis, Ferenc Preisach introduced the relay hysteron as an elementary mathematical description. In this new method (SINDyHybrid), tailored basis functions in form of relay hysterons are added to the library which is used by SINDy. Experiments with a hysteresis controlled water basin show that this approach correctly identifies state transitions of hybrid systems and also succeeds in modeling the dynamics of the discrete system states. A novel proximity hysteron achieves the robustness of this method. The impacts of the sampling rate and the signal noise ratio of the measurement data are examined accordingly.
\end{abstract}


\section{Introduction}
Nowadays, data-driven approaches for modeling dynamics without a fixed model structure gain relevance. However, classical grey-box modeling is still popular because it allows to achieve transparency and to include domain-specific know\-ledge as well as constraints. The presented approach allows to combine explicit knowledge with a flexible data-driven approach in order to estimate hysteresis behavior.

Whenever a certain degree of precision is required for tasks like optimization or anomaly detection, simple dynamic models may deviate too far from the actual system, especially in the critical regions of hysteretic systems, where the state transitions happen. In order to pursue model-based approaches for these tasks, sophisticated models are required.

A traditional approach of obtaining a system model combines physical laws and technical parameters to so called white-box-models. Alternatively, grey-box-models can be used, which have a fixed model structure but chosen parameters are tuned using measurement data. Methods from machine learning partially abandon fixed model structures, like in the case of neural networks. These approaches do not provide transparency,  which implies a need for empirical validation. Furthermore, pure data-driven methods require for extensive data for all considered situations and operation points. However, due to high costs, efforts or dangers it is not always feasible to conduct real experiments. SINDyHybrid on the other hand does not demand a fixed model structure like grey-box-models. Not only the parameters, but also the model structure is deduced from measurement data, using a sparse regression method. While not as flexible as neural networks, this approach leads to easily readable and transparent models.

This article proposes an extension of the Sparse Identification of Nonlinear Dynamics (SINDy) framework developed and presented by S. Brunton and J.N. Kutz \cite{BruntonSINDy.2016}. Adding the robust identification of hysteresis behavior in hybrid systems holds the potential to identify all discrete states of a hysteresis controlled system and to build a nonlinear model for the dynamics on all states in one step. The approach aims to handle diverse systems, to include system knowledge and to result in an easily interpretable model.

Sec. \ref{sec:stateOfTheArt} covers basic theory regarding system identification and current approaches. A new concept is presented in Sec. \ref{sec:Concept}. Sec. \ref{sec:practicalEvaluation} contains the practical experiments and their results while Sec. \ref{sec:discussion} gives a deepening discussion. Sec. \ref{sec:conclusion} summarizes the findings and gives a brief outlook to further applications.

\section{System Identification for nonlinear hybrid systems}	
\label{sec:stateOfTheArt}

Firstly, the considered dynamic systems are hybrid, e.g. state-space is augmented by a discrete variable which denotes different continuous dynamics. Vice versa, the continuous states of the system can lead to switches in the discrete variable. For example the behavior of water changes with reaching the threshold of 100 $^\circ C$ due to the phase transition. Therefore, it makes sense to create two distinct models: one for the discrete state \textit{temperature below 100 $^\circ C$} and one for \textit{temperature above 100 $^\circ C$}.

Hysteresis behavior evokes a special type of hybrid systems which requires more complex modeling methods. Sec. \ref{sec:modelingHybridSystems} discusses established methods for modeling hybrid systems and their respective limitations. The ambition to model hysteresis behavior motivates to explain the fundamentals of Koopmanism and its application in SINDy (Sec. \ref{sec:koopman}).

\subsection{Hysteresis Modeling}
\label{hysteresisModeling}

Whenever a system's behavior does not only depend on its current internal states, but also on its past trajectory, the system is called \textit{hysteretic}. This nonlinear effect occurs in natural phenomena like in magnetization of ferromagnetic materials, but also in artificial systems such as two-point controllers. The Hungarian physicist Ferenc Preisach presented the first suitable model to describe hysteretic behavior of magnetism in 1935\cite{Preisach.1935, PreisachInMemorial}. His hysteron-operator allows for approximation of natural hysteresis using weighted stairs. The respective relay operator $R_{\alpha, \beta}$, also called relay hysteron, is expressed by
	\begin{equation}\label{eq:relayHysteron}
	y(x,t) = \begin{cases}
	1, & \text{if $x(t) \geq \beta$} \\
	0, & \text{if $x(t) \leq \alpha$} \\
	y_p, & \text{if $ \alpha < x(t) < \beta $,}
	\end{cases}
	\end{equation}	

\begin{figure}
\centering
\includegraphics[width=0.82\linewidth]{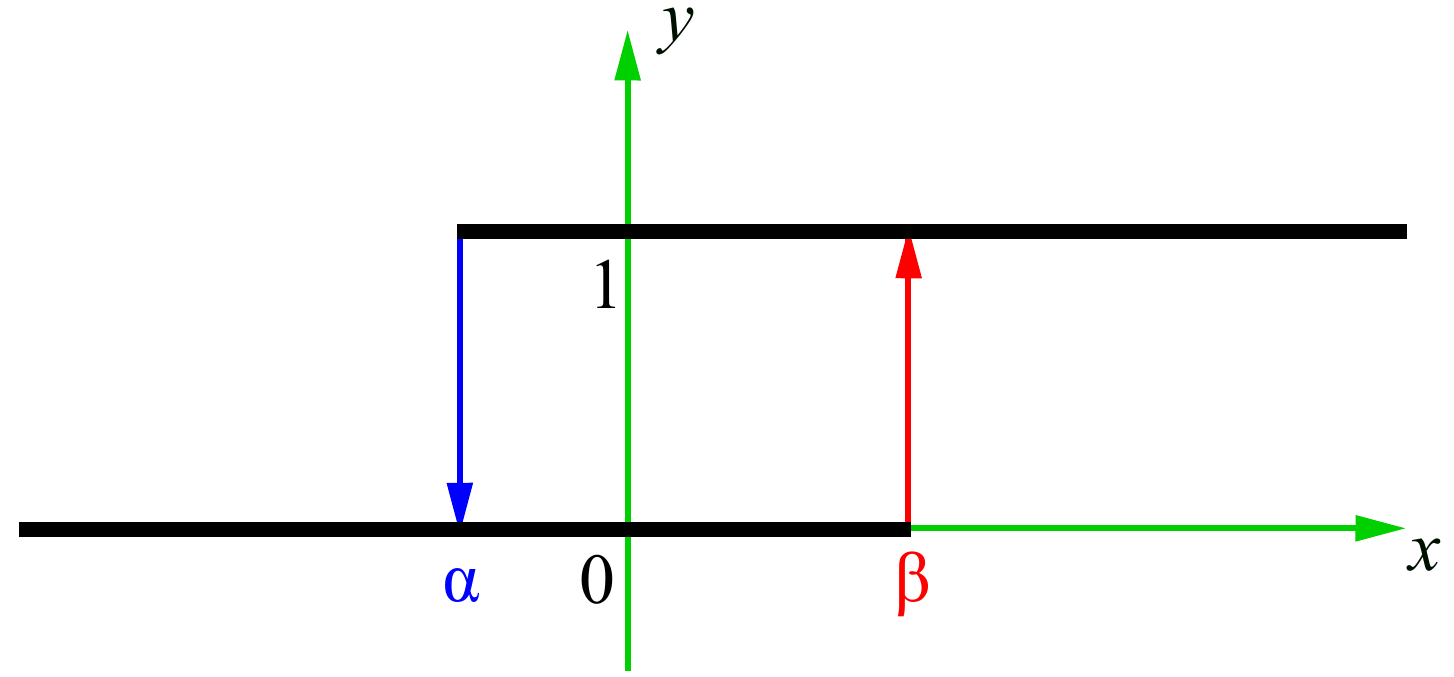}
\caption{Relay hysteron \cite{RelayHyst1}}
\label{fig:preisachRelay}
\end{figure}

where $y_p$ is $0$ if $x(t)\leq \alpha$ is more recent. Vice versa, $y(x,t)$ remains $1$ if $x(t)$ comes from $x(t)\geq \beta$, see Fig.\ref{fig:preisachRelay}. The complex shape of a magnetic hysteresis curve can be achieved by superposition of many elementary relay hysterons. Modern approaches of hysteresis modeling are discussed in\cite{Hassani.2014}. We utilize the concept of hysterons which is the basis of Preisach's model.

\subsection{Modeling of hybrid systems}
\label{sec:modelingHybridSystems}

\changepage{-27pt}{}{}{}{}{27pt}{}{}{}

Modeling of hybrid systems demands for model structures which contain both discrete and continuous states. Subsequently, established types are presented.

\subsubsection{PWARX models}
\label{PWARX} 

Piecewise-affine (PWA) methods divide the state-space into discrete regions, commonly shaped as polyhedrons \cite{Ferrari-Trecate.2000}. For every region a continuous model is set up, e.g. the well known linear autoregressive model with exogenous input (ARX-model). This procedure is called PWARX\cite{pwarx_paoletti}. If the transition from one state to another is affected by hysteresis behavior, PWA models are not sufficient anymore, because they assume fixed transition borders from one discrete state to another. They are unable to include memory of the past trajectory of the system's state.

\subsubsection{SARX models} \label{SARX} Switched affine autoregressive models with exogenous input (SARX) can be used if transitions of the state are triggered by external events. It is also possible to use piecewise continuously differentiable nonlinear functions (PWNARX, SNARX). An overview of related techniques is given in \cite{Hutchison.2008}. Similar to PWARX, SARX models with their fixed transitions fail to model hysteretic behavior because they do not consider any system memory.

\subsubsection{Hammerstein models}
In 2015, Fang et al. presented a data-driven approach to model sticky valves using the Preisach model in form of a Hammerstein model \cite{Fang.2015}. A Hammerstein system consists of two coupled blocks: a static nonlinearity and a linear time-invariant system, e.g. the above explained ARX models. The authors propose an extended version which involves the memory-effect of the hysteresis. They succeed in modeling the state transitions, but they separate the identification of the transition model from the identification of dynamic models for the discrete states. Also, the number of discrete states must be known beforehand. This will be overcome by using SINDyHybrid.

\subsubsection{Takagi-Sugeno fuzzy model}

Piezoelectric actuators (PEA) show a magnetic hysteresis behavior. In \cite{LongCheng2017AATF}, a Takagi-Sugeno fuzzy model predictive controller utilizes a nonlinear autoregressive moving average with exogenous inputs (NARMAX) model. A fuzzy inference module concludes the control output, based on several local MPC-rules. Similar to the Hammerstein-approach above, the structure is customized to the considered natural hysteresis. In contrast, SINDyHybrid promises a broader applicability for hybrid systems.

\subsubsection{Black-box-techniques}
An alternative approach would be to use neural networks to learn the system dynamics, but the resulting models are usually difficult to check for plausibility and stability, which poses further problems for industrial applications \cite{Ljung_BlackBox_2001}.

\subsection{From Koopman to system identification}
\label{sec:koopman}
This section will briefly introduce the Koopman operator and the concept of sparsity, as well as present an overview of the SINDy methodology and its applications. In the concluding Sec. \ref{sec:Concept}, 
a concept is shown which implements the hysteresis model into the given framework.
\subsubsection{Koopman operator}
The Koopman operator $\mathcal{K}_t$\cite{Koopman.1931} is a linear but typically infinite-dimensional operator, which is capable of describing the full dynamics of an underlying system. For this paper we focus on the practical implementation of Koopman operator theory. An in-depth overview about theoretical concepts and applications can be found in \cite{Budisic.2012}. In \cite{bakker2020learning}, three Koopman representations for hybrid systems are presented, two of which eliminate the discrete state variables entirely.

Consider a dynamical system of the form \mbox{$\dot{x} = f(x)$} with state $x \in \mathbb{R}^n$ and flow $F^t$ defined by $F^t(x(t_0)) = x(t_0+t)$. Now consider real valued observable functions of the state, $h(x)$, as elements of some infinite-dimensional Hilbert space. The Koopman Operator $\mathcal{K}_t$ is then defined by
\begin{equation}
\left(\mathcal{K}^t h\right)(x) = h\left(F^t(x)\right).
\label{eq:koopman}
\end{equation}

In the continuous case one often considers the infinitesimal generator $\mathcal{K}$ of the operator family $\{\mathcal{K}_t\}_t$, which satisfies

\begin{equation}
\dfrac{d}{dt}h(x) = \mathcal{K}h(x).
\label{eq:infinitesimal_generator}
\end{equation}

Hence, $\mathcal{K}$ evolves observable functions  $h(x)$ instead of direct traces of the states. These can be arbitrary functions, i.e.:
\begin{equation}
\label{eq:observables_example}
h(x) = 1, |x|^2, \sin(x_1), e^{x_n}, \hdots
\end{equation}
Any finite subset of such observable functions span a subspace of the Hilbert space \cite{BruntonKoopmanInvariant.2016b}.
In general, $\mathcal{K}$ in Eq. \eqref{eq:infinitesimal_generator} is infinite dimensional but linear. This enables tools like spectral analysis without any inaccuracy caused by linearization \cite{Budisic.2012}. In 2005, Igor Mezi\'{c} presented the application of the Koopman operator for spectral analysis of high-dimensional, nonlinear systems \cite{Mezic.2005}. Presented by Schmidt et al. in 2010, the Dynamic Mode Decomposition (DMD) extracts eigenvalues and modes of the Koopman operator from measurement data\cite{Schmid2010}, i.e. it finds the most probable system matrix $A$ for \mbox{$X_{k+1} \approx A X_k$.} The extended-DMD (e-DMD) additionally approximates eigenfunctions of the Koopman operator\cite{WilliamsKernel,WilliamsEDMD.2015}.

\subsubsection{Sparsity}
The concept of sparsity is useful to find solutions of underdetermined systems of linear equations. A solution is considered \emph{sparse}, when most of its entries vanish\cite{Cands2014MathematicsOS}. Next to offering numerical advantages, a solution with a limited number of terms facilitates the manual interpretation of the results. The problem of finding sparse solutions can be connected to model reduction techniques, as discussed in \cite{loiseau_brunton_2018}.

In the context of sparse sensing, Donoho showed that an underdetermined system of equations can be converted to a convex problem using a sparsity promoting condition \cite{Donoho.2006}. Consider solving
\begin{equation}
	y = \theta s^{\prime}
\end{equation}
for $s^{\prime}$, with $y \in \mathbb{R}^n$, $\theta \in \mathbb{R}^{n \times m}$, $s^{\prime}   \in \mathbb{R}^m$ and $m>n$. The parameter vector $ s^{\prime}$ needs to be determined in order to fit the measurement vector $y$ using the product of the observation matrix $\theta$ and $ s^{\prime}$.

The common $\text{L}_2$-regularized regression promotes a high number of involved terms. On the contrary, by using the $\text{L}_1$-norm as regularization term, terms of minor impact are neglected from the solution. The $\text{L}_1$-regularized regression 
\begin{equation}\label{eq:l1regression}
s = \argmin_{s^{\prime}} \| \theta s^{\prime} - y \|_2 + \lambda \| s^{\prime} \|_1
\end{equation}
 with weight $\lambda$ can be solved e.g. by the LASSO method \cite{TibshiraniLasso}.
The solution $s$ is optimal in $\text{L}_1$-sense and sparse within the space spanned by the columns of $\theta$, i.e. the basis functions. This concept of sparsity is, along with Koopman theory, the basis for SINDy.

\subsubsection{Sparse Identification of Nonlinear Dynamics (SINDy)}		\label{sec:sindy}
Since 2015, Nathan Kutz, Steven Brunton and Joshua Proctor have developed the SINDy method\cite{BruntonSINDy.2016}, which is the basis for our data-driven approach to identify hybrid behavior. First, a library $\Theta(X)$ with possible basis functions is built. SINDy aims to find the minimal amount of those functions, with which a given signal can be approximated.

In contrast to DMD, which results in a linear model, SINDy can include arbitrary nonlinearities. As a further development, SINDYc can include feedback and control\cite{BRUNTON_SINDYc}. A successful application of SINDy for predictive control is reported in \cite{Kaiser.}.

The main principle of SINDy is the approximation of signals using only few basis functions, i.e. \textit{sparse} solutions. The vector
$x(t) = \begin{bmatrix}
x_1(t) &	x_2(t)	& \dots & x_n(t)
\end{bmatrix}^T \in \mathbb{R}^n $
represents all states of the system at time $t$. Measurement data of these states $x(t)$ and their derivatives $\dot{x}(t)$ at the time-steps $t_1, t_2, \dots , t_m$ form two data matrices $X$ and $\dot{X}$, given by
\begin{align}
\label{eq:dataMatricesX}
	X &= \begin{bmatrix}
		x^T(t_1) \\ 
		x^T(t_2) \\
		\vdots \\
		x^T(t_m)
	\end{bmatrix}
	=	\newcommand{\mymatrix}[1]{\ensuremath{\overset{\xrightarrow[\hphantom{#1}]{\text{\scriptsize States}}}{#1}\left\downarrow\vphantom{#1}\right.}}
	\mymatrix{\begin{bmatrix}
		x_1(t_1)	&	x_2(t_1)	& \dots	&	x_n(t_1) \\
		x_1(t_2)	&	x_2(t_2)	& \dots	&	x_n(t_2) \\
		\vdots		&	\vdots		& \ddots&	\vdots	 \\
		x_1(t_m)	&	x_2(t_m)	& \dots	&	x_n(t_m) \\
		\end{bmatrix}}
		\text{\scriptsize Time}
\\
\dot{X} &=
\begin{bmatrix}
\dot{x}^T (t_1) \\ 
\dot{x}^T (t_2) \\
\vdots \\
\dot{x}^T (t_m)
\end{bmatrix}
= \begin{bmatrix}
	\dot{x}_1(t_1)	&	\dot{x}_2(t_1)	& \dots	&	\dot{x}_n(t_1) \\
	\dot{x}_1(t_2)	&	\dot{x}_2(t_2)	& \dots	&	\dot{x}_n(t_2) \\
	\vdots		&	\vdots		& \ddots&	\vdots	 \\
	\dot{x}_1(t_m)	&	\dot{x}_2(t_m)	& \dots	&	\dot{x}_n(t_m) 
	\end{bmatrix} .
	\label{eq:dataMatricesXdot}
\end{align}

As a next step, a library  $\Theta(X)$  is built from candidate functions, in which $X$ is evaluated. Typical candidates are polynomials, trigonometric functions and constants, but arbitrary functions can be included as well, like in
\begin{align}
\Theta(X)\!= \!
\begin{bmatrix}
\vline 	&\vline & \vline 	&\vline 	& 		&	\vline 	& \vline  & 		& \\
1 		&	X 	& X^{P_2} 	& X^{P_3} 	& \!\dots\! & \sin(X) 	& \cos(X) & \!\dots\! 	& e^X  \\
\vline 	&\vline & \vline 	&\vline 	& 		&	\vline 	& \vline  & 		&
\end{bmatrix}.
\label{eq:bibliothekStruktur}
\end{align}		

The terms $ X^{P_n} $  are polynomials of degree $n$. This includes cross terms of the single states. For the case of quadratic polynomials, this yields
\begin{equation}
\delimitershortfall=0pt
\begin{footnotesize}
	X^{P_2} \!= \!
\begin{bmatrix}
x_1^2(t_1) \!&\! x_1(t_1)x_2(t_1) \!& \!\dots\!	&\! x_2^2(t_1) &\! x_2(t_1)x_3(t_1) & \!\dots\! &\! x_n^2(t_1) \\
x_1^2(t_2) \!&\! x_1(t_2)x_2(t_2) \!& \!\dots\! &\! x_2^2(t_2) &\! x_2(t_2)x_3(t_2) & \!\dots\! &\! x_n^2(t_2) \\
\vdots	   \!&\! \vdots			  \!& \!\ddots\!&\! \vdots	   &\!	\vdots		  & \!\ddots\!&\!\vdots     \\
x_1^2(t_m) \!&\! x_1(t_m)x_2(t_m) \!& \!\dots\!	&\! x_2^2(t_m) &\! x_2(t_m)x_3(t_m) & \!\dots\! &\! x_n^2(t_m)
	\end{bmatrix} . \label{eq:polynomialLib}
\end{footnotesize}
\end{equation}

So, the dynamics can be expressed by $\dot{X} = \Theta(X) \Xi$, where $\Theta(X)$ is the evaluated library and $\Xi = \begin{bmatrix}
\xi_1 & \xi_2 & \dots & \xi_n
\end{bmatrix}$ is the coefficient matrix. The coefficients can be determined using sparse regression techniques \cite{BruntonSINDy.2016}.

For time-discrete systems, the equation
\begin{equation}\label{eq:sindyDiscrete}
X_{k+1} = \Theta(X_k) \Xi
\end{equation}
can be solved accordingly. SINDy allows for easy identification of few dominant functions out of a basis library. Choosing an appropriate library, which contains sufficient functions to describe the dynamics, is essential for a successful identification.
\section{Concept for identification of hysteresis-controlled systems}
\label{sec:Concept}
This section provides a technical concept for system identification of hysteretic systems using SINDy, named SINDyHybrid. Based upon theoretical considerations of SINDy and hybrid systems, we developed an approach to incorporate tailored basis functions into the function library. 

SINDy has been used by Brunton et al. to model the dynamics of systems with a bifurcation parameter $\mu$, which changes the qualitative behavior of the system's dynamics \cite{BruntonSINDy.2016}. In order to incorporate $\mu$ into the observables, the equation $\dot{x} = f(x)$ was extended by $\mu$, such that
\begin{align}
	\dot{x} &= f(x, \mu) \label{eq:sindyBifurcation}\\
	\dot{\mu} &= 0. \label{eq:MuDerivative}
\end{align}
The function library $\Theta$ then consists of functions of $x$ and $\mu$. The value of $\mu$ however remains unchanged, as indicated by Eq. \eqref{eq:MuDerivative}. A more general formulation has been used in \cite{BRUNTON_SINDYc}, where feedback control signals were included by choosing 
\begin{equation}
	\dot{\mu} = g(x). \label{eq:MuDerivativeSindyC}
\end{equation}
Eq. \eqref{eq:MuDerivativeSindyC} expresses that changes of the bifurcation parameter $\mu$ depend uniquely on the continuous state $x$ and not on its own past. So, $\mu$ is not a separate state but an observation of $x$.
\subsection{Expanding the theory for hybrid systems}
We now further generalize the approach of Eq. \eqref{eq:MuDerivativeSindyC} by choosing 
\begin{equation}
	\dot{\mu} = g(x, \mu), \label{eq:MuDerivativeGeneral}
\end{equation}
which connects changes of $\mu$ to current values of $\mu$. By adding $\mu$ as an argument to the function, state transitions depend on both the continuous state $x$ and the discrete system state $\mu$.

Given a state $x$ and using Eq. \eqref{eq:MuDerivativeSindyC}, the system will always evolve in the same way. Adding $\mu$ as an argument to $g$ as in Eq. \eqref{eq:MuDerivativeGeneral} allows for different trajectories from the same starting point $x$, depending on its discrete state $\mu$. This represents the behavior of a hybrid system. Special cases of Eq. \eqref{eq:MuDerivativeGeneral} can be linked to the previously presented methods of modeling hybrid systems. For example, setting an explicit time dependency
\begin{equation}
	\dot{\mu} = g(t)
\end{equation}
results in the formulation of SARX models (see Sec. \ref{SARX}). In cases where the change of the qualitative behavior of the system depends only on current values of the state $x$, the change of $\mu$ can be represented by 
\begin{equation}
	\dot{\mu} = g(x(t)), \label{eq:MuDerivativePWARX}
\end{equation}
which is the basic assumption for piecewise-affine (PWA) models (see Sec. \ref{PWARX}). For a linear function $g$, the regions for each state have a linear border. For a nonlinear $g$, complex borders can be modeled between regions, as is the case for NPWA models. If the function $f$ in Eq. \eqref{eq:sindyBifurcation} is linear, the formulation is equivalent to PWARX models, otherwise it results in PWNARX models.

Hysteresis models not only depend on the current state of the system, but also on past values up to a horizon $q$. For setting $\mu$ dynamically depending from past values, we define
\begin{equation}
	\dot{\mu} = g\left(x(t-q), x(t-q+1), ..., x(t-1), x(t), \mu\right) = g\left(x_q, \mu\right).
\end{equation}
This way, $\mu(t)$ serves as a storage for information of past values of the system. For a technical (rectangular) hysteresis, considering only the present value of $\mu$ is sufficient, leading to \eqref{eq:MuDerivativeGeneral}.

\subsection{Modeling discrete states}\label{modelingDiscreteStates}
In order to find a suited function $g$ from Eq. \eqref{eq:MuDerivativeGeneral} to model changes of discrete systems, one has to consider the way the change of states takes place. The transition can for example be either abrupt or smooth, following a certain shape. In the case of two-point controlled systems the state transition is abrupt: As soon as a certain threshold is passed, the system behavior changes more or less instantaneously.

In \ref{hysteresisModeling} we discussed the Preisach model for identification of hysteresis, which uses the relay operator $R_{\alpha, \beta}$. Based upon the idea that state transitions in many hybrid systems happen abruptly, the relay hysteron seems to be a suited candidate for the function $g$. %In this way, hysteresis behavior can be included in the customized library for SINDy.

\subsection{Preparing the data}	
				
The first step of data processing is affine scaling of all data to a sensible and comparable range, e.g. from -1 to 1. This promotes an equal treatment of all time series, not depending on the absolute values of the data. 
	
Before identification with SINDy begins, the possible state indicators (relay hysterons, Sec. \ref{modelingDiscreteStates}) have to be computed from the data. This is done in three steps: Building indicator functions, pairwise matching of indicator functions and evaluating the relay hysterons.
	
We assume a change of the system behavior as soon as a value $x$ rises above $\beta \in \mathbb{R} $ or falls below $\alpha \in \mathbb{R} $. We also assume $\alpha$ and $\beta$ are included in our data. The first step is to build differences between all time series in our data. If the units of the time series are known, it is helpful only to build differences between commensurable data, e.g. time series which share the same unit, as exemplarily done in Eq. \eqref{eq:differences}.
\begin{equation}
	\begin{aligned}
		\label{eq:differences}
		\tilde{x}_1 &= x-\alpha \\
		\tilde{x}_2 &= x-\beta \\
		\tilde{x}_3 &= \alpha - \beta
	\end{aligned}
\end{equation}
We then create indicator functions,
\begin{align}
\begin{split}
I(x) &= \begin{cases}
1,  & \text{if $x \geq 0 $} \\
0,  & \text{if $x < 0 $} \\
\end{cases} \label{eq:indicator1} \\
\bar{I}(x) &= \begin{cases} 
1,  & \text{if $x < 0 $} \\
0,  & \text{if $x \geq 0 $}, \\
\end{cases}
\end{split}
\end{align}
masking these differences \eqref{eq:differences} depending on the sign of the value. For the upper threshold, $I_\beta = I(\tilde{x}_2)$ masks ranges, where $x \geq \beta$. For the lower threshold, $I_\alpha = \bar{I}(\tilde{x}_1)$ masks ranges, where  $x < \alpha$. 

A hysteron can be formed out of two indicator functions which are never both true at the same time. The hysteron $H_{\alpha, \beta}$ is built by pairing two indicators $I_\alpha, I_\beta$. This way,  $I(\tilde{x}_2)$ and $\bar{I}(\tilde{x}_1)$  are also combined. For each hysteron, a complementary hysteron $\bar{H}_{\alpha, \beta}$ is built. The evaluation is done according to

\begin{equation}\label{eq:hysteron}
H(k)_{\alpha, \beta} = \begin{cases}
1, & \text{if $I_\beta = 1$.} \\
0, & \text{if $I_\alpha = 1$.} \\
H(k-1), & \text{for $I_\beta = I_\alpha = 0$.}
\end{cases}
\end{equation}
\begin{center}
	for each step $k>1$. 
\end{center}
	
This is equivalent to the formulation of the relay hysteron in Eq. \eqref{eq:relayHysteron}. Presuming alternating switches between the states, $H(1)_{\alpha, \beta}$ can be initialized with the opposite state of the first occurring switch. This presumption has to be reviewed depending on the use case.

\subsection{Robust handling of transitions -- the proximity hysteron}
\label{sec:proximity}
A problem in the design of hysterons is the fact that the critical threshold values $\alpha$ and $\beta$ from Eq. \eqref{eq:hysteron} are only reached for a short moment. Due to a slow sampling frequency or noisy data it may happen that these critical points are not even included in the data at all. In order to improve robustness of the relay hysterons, we developed a \textit{proximity-hysteron}, which utilizes an $\epsilon$-range around the actual threshold value for switches between states. The evaluation is done according to 

\begin{equation}\label{eq:proxHysteron}
H_{\epsilon}(k)_{\alpha, \beta} = \begin{cases}
1, & \text{if $I_{\epsilon, \beta}  = 1$,} \\
0, & \text{if $I_{\epsilon, \alpha} = 1$,} \\
H_{\epsilon}(k-1), & \text{for $I_{\epsilon, \beta} = I_{\epsilon, \alpha} = 0$,}
\end{cases}
\end{equation}

with the proximity indicator functions %$I_{\epsilon_\beta}(x)$

\begin{align}
\begin{split}
I_{\epsilon_\beta}(x) &= \begin{cases}
1,  & \text{if $x = \min\{\max( Z_{x_\beta}(t) ), \beta\} $,} \\
0,  & \text{otherwise,} 
\end{cases} \\
I_{\epsilon_\alpha}(x) &= \begin{cases}
1,  & \text{if $x = \max\{\min( Z_{x_\alpha}(t) ), \alpha\} $,} \\
0,  & \text{otherwise,} 
\end{cases}
\end{split}
\end{align}

and the connected sets \mbox{$Z_{x_\beta}(t) = \{x \mid x \geq \beta - \epsilon_\beta\}$} and \\ $Z_{x_\alpha}(t) = \{x \mid x < \alpha + \epsilon_\alpha\}$, with $\epsilon_\beta,\epsilon_\alpha \in \mathbb{R} > 0$, which include $x(t)$.

\begin{figure}
	\begin{center}
		\includegraphics[width=\linewidth]{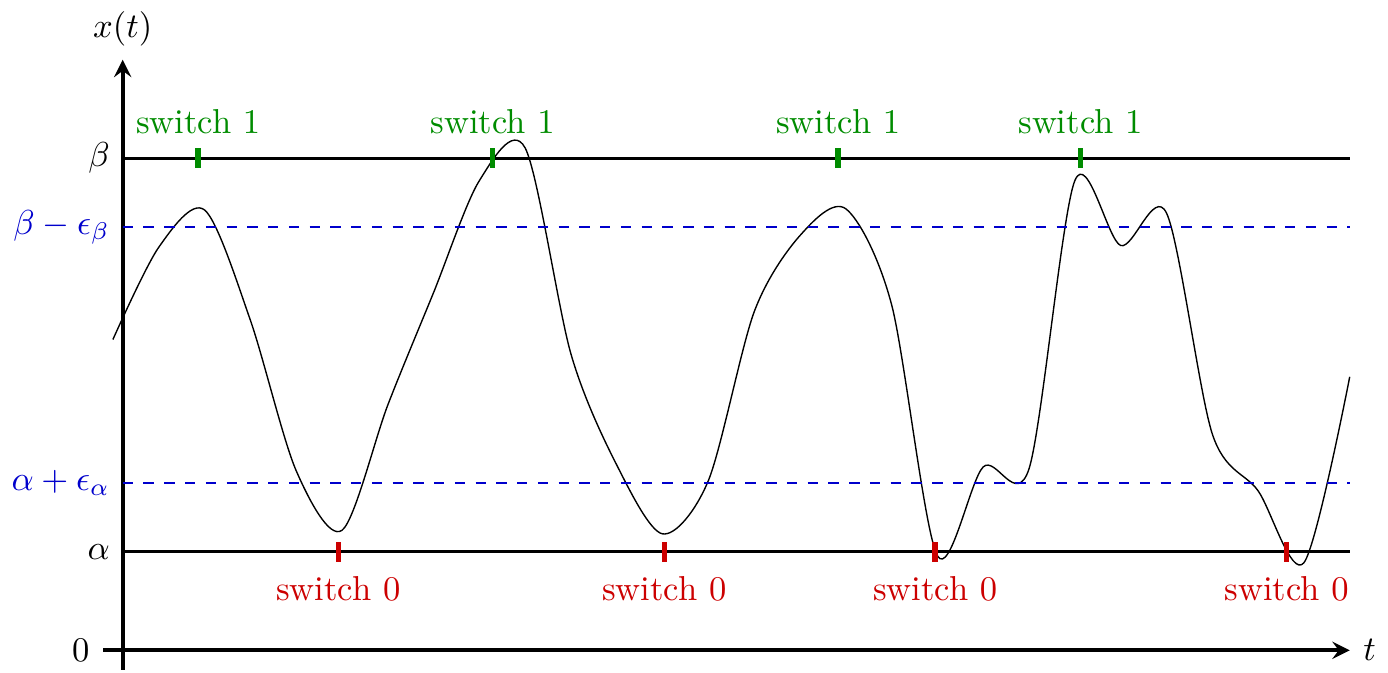}    % The printed column width is 8.4 cm.
		\caption{Switches of the proximity hysteron} 
		\label{fig:proximity_hysteron}
	\end{center}
\end{figure}

The proximity indicator functions define an $\epsilon$-range, which moves the threshold for the transition in the corresponding direction (see Fig. \ref{fig:proximity_hysteron}). However, the transition between states does not happen as soon as the shifted threshold is reached, but only when the extremum within the connected set of points within the $\epsilon$-range is reached. An exception occurs, when the actual threshold $\alpha$ or $\beta$ is reached. In this case, the switch is done regardless of whether this is the extremum within this set.

\begin{figure}
	\begin{center}
		\includegraphics[width=\linewidth]{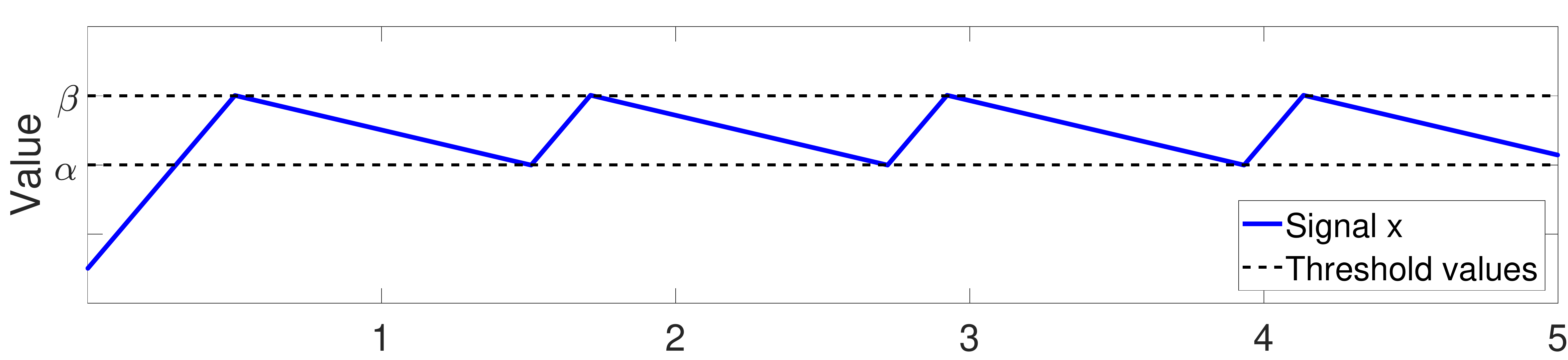} \\ 
		\vspace{-0.3mm}  
		\includegraphics[width=\linewidth]{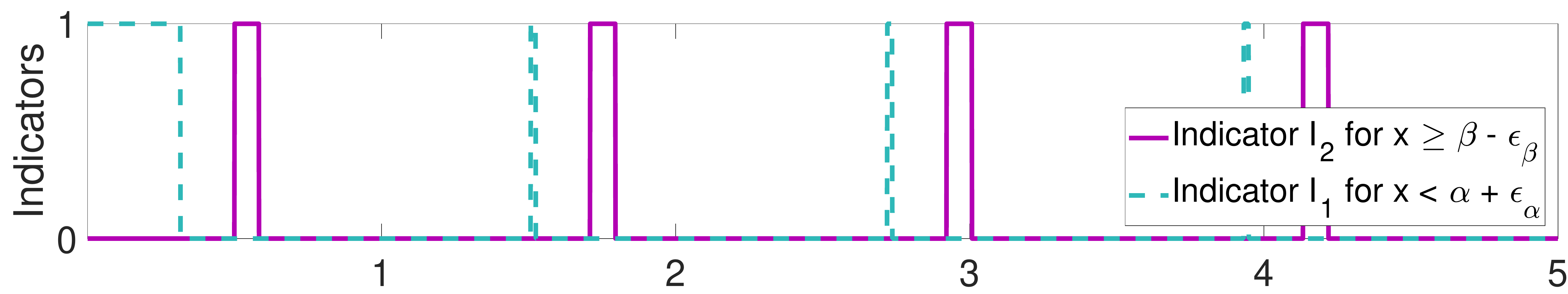} \\
		\vspace{-0.3mm}
		\includegraphics[width=\linewidth]{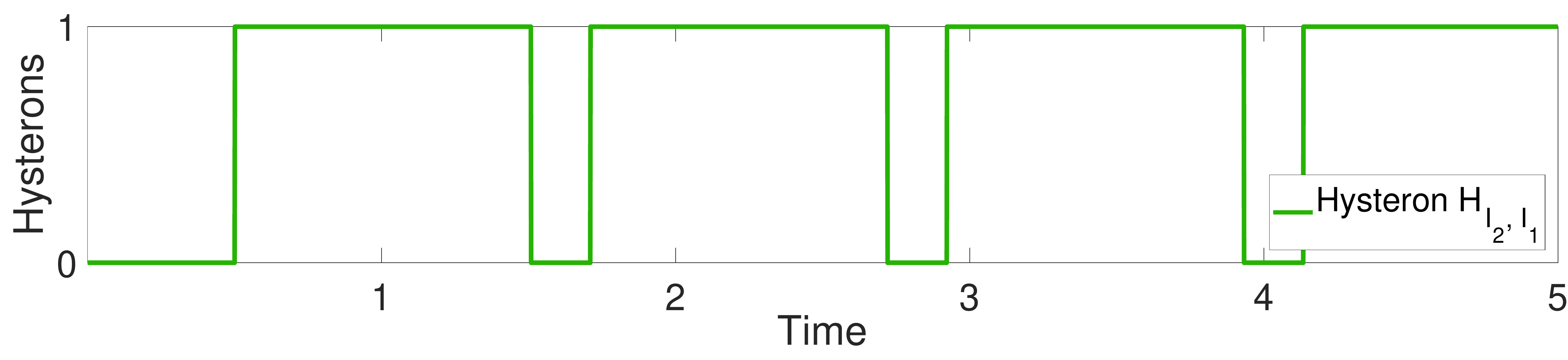} 
		\caption{Interaction between signal, threshold values, indicator functions and hysterons} 
		\label{fig:buildingOfHysterons}
	\end{center}
\end{figure}

Due to the introduction of the proximity hysteron, state transitions can even be localized under harsh conditions -- if the threshold points do not occur in the measurement data. The connection between signal, threshold values, indicator functions and hysterons is visualized in Fig. \ref{fig:buildingOfHysterons}.

%% \subsubsection{Extending the state space} - Possibly leave out!
	
\subsection{System Identification with SINDy}
\begin{figure}
	\begin{center}
		\hspace{-1.4cm}
		\includegraphics[width=0.9\linewidth]{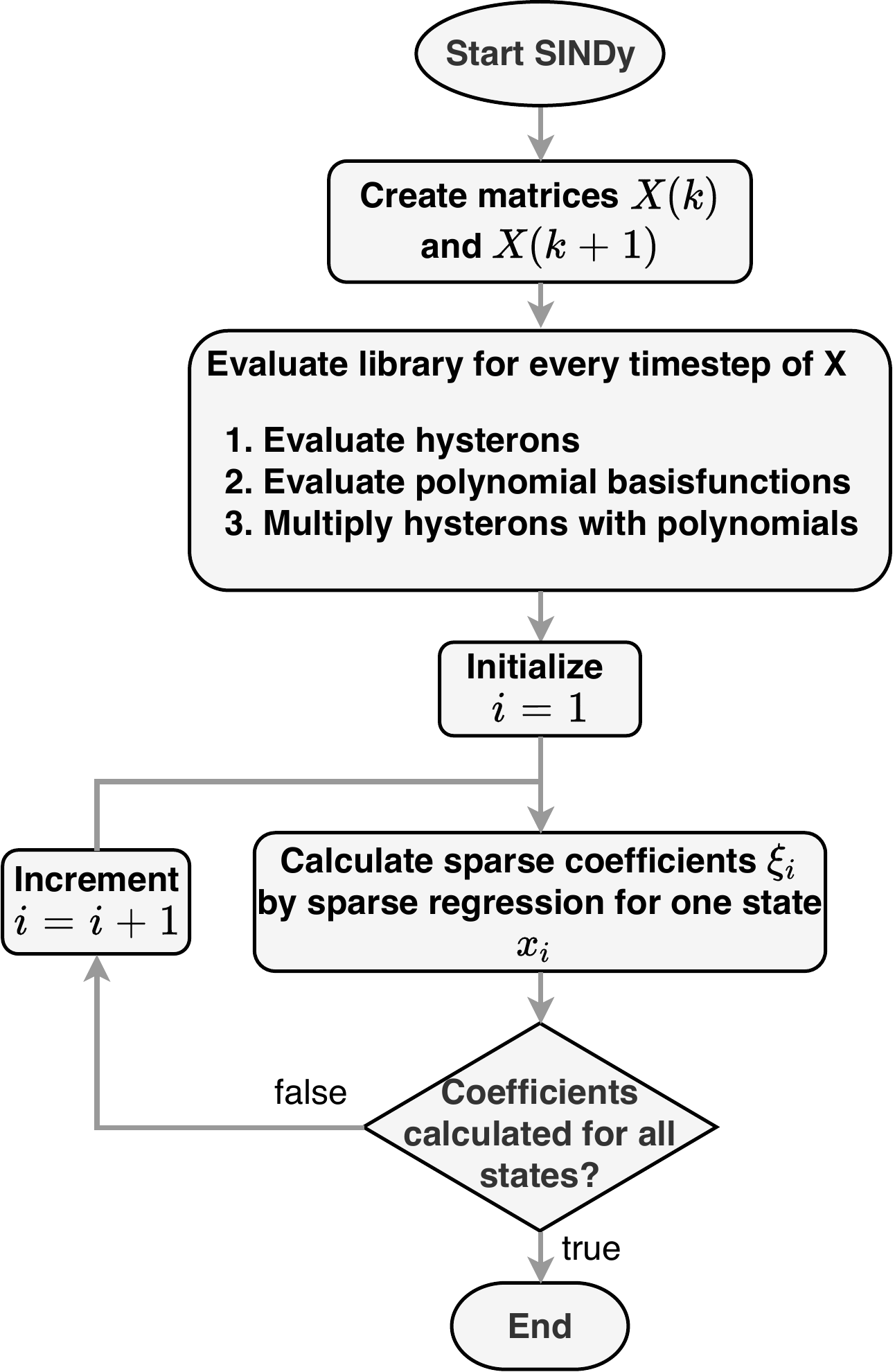}    % The printed column width is 8.4 cm.
		\caption{The SINDy algorithm with tailored basis functions for identification of hysteretic systems} 
		\label{fig:sindyFlow}
	\end{center}
\end{figure}

Once the hysterons are built and other preprocessing measures have been taken, the SINDy algorithm can start. The schematics are  shown in Fig. \ref{fig:sindyFlow}. Creating the data matrices includes choosing the right state vector $X$. The past values $H(k-1)_i$ for $i = 1, ..., m$ (see Eq. \eqref{eq:proxHysteron}) of all $m$ hysterons have to be included, holding information about the system's current discrete state. Past measurement values $x(k-1), ..., x(k-q) $ up to a defined horizon $q$ can also be considered, allowing a representation resembling a filter with finite impulse response (FIR). The state at time $k$ can be expressed by

\begin{align}
\begin{split}
	X(k) =& [x_1(k), ... , x_n(k), H_1(k-1), \bar{H}_1(k-1), ...
	\\& \bar{H}_m(k-1), x_1(k-1), ... , \bar{H}_m(k-q)].
\end{split}
\end{align}

For the design of the library $\Theta$, polynomial basis functions are useful and easy to include. The function $\varphi_\text{poly}(x_1, x_2, ..., x_n)$ evaluates all polynomials between $x_1$ to $x_n$ up to a predefined polynomial degree. Hysterons are evolved by $\varphi_\text{relay}(x_1, x_2, ..., x_n, H_i)$. After evaluating all basis functions, all those basis functions that are unaffected by hysterons are multiplied with the updated hysterons. The finished library $\Theta$ consists of three parts: One without hysterons, the second with cross terms of hysterons and non hysteretic terms and the third of only the updated hysterons. Once the library is evaluated, a sequential least squares regression with a tuning parameter $\lambda$ is used to solve $ X(k+1) = \Theta\left(X(k)\right) \Xi$ for the sparse coefficients $\Xi$, as proposed in \cite{BruntonSINDy.2016}. 

\begin{figure}
	\begin{center}
		\includegraphics[width=\linewidth]{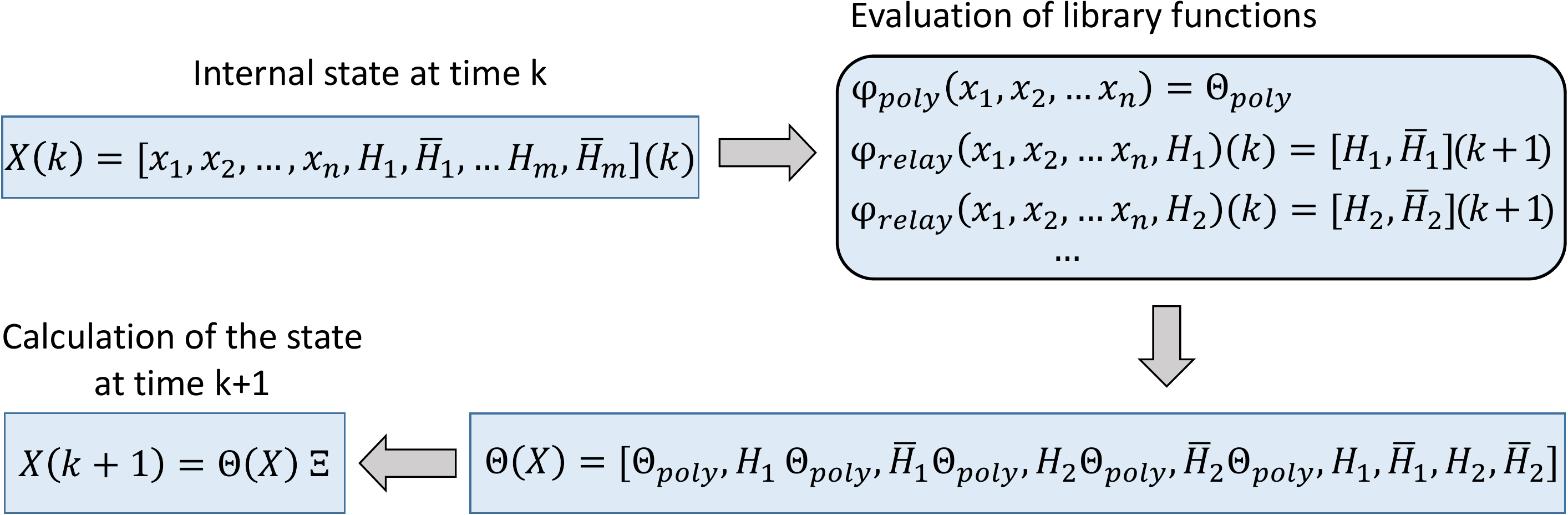}    % The printed column width is 8.4 cm.
		\caption{Building and evaluation of the library}
		\label{fig:sindyLib}
	\end{center}
\end{figure}

For a library consisting of only polynomial basis functions and tailored basis functions for the propagation of hysterons, the evaluation and calculation of the next step is shown in Fig. \ref{fig:sindyLib}.
 Both, the sparse coefficients $\Xi$ and the prepared library $\Theta$ form the actual result of the SINDy method, the dynamic model of the observed system. First experiments with this new approach are documented in the following section.
\section{Practical evaluation and results}
\label{sec:practicalEvaluation}
In order to evaluate the presented concept, we used different model parameters and differing data quality to identify a simple system. Our aim was to determine 
\begin{enumerate}
\item under which conditions the proposed identification procedure succeeds in finding a suitable model,
\item which parameters influence model quality,
\item and which aspects limit this approach.
\end{enumerate}
 The influence of the size of the library is studied in the first experiment by varying the degree of polynomials used. The robustness to sample rate and measurement noise are further aspects under test. 
\subsection{Example: Hysteresis-controlled tank system}		
The level of the basin in Fig. \ref{fig:Wassertank} is controlled by a \textit{two-point controller} with upper set point  $h_{\text{max}}$ and lower set point $h_{\text{min}}$. A leak causes a constant drain $Q_{\text{out}} \in \mathbb{R} < 0$, which neglects the influence of the current water level. The control signal $u(t)$ turns the pump on whenever $h_{\text{min}}$ is reached, resulting in a steady inflow $Q_{\text{in}} \in \mathbb{R} > 0$. The pump is turned off when the level raises over $h_{\text{max}}$. The length of the supply pipe $l_\text{P}$ can be used to analyze the effect of delay.
\begin{minipage}{0.5 \textwidth}
	\input{tank_model}

	\label{tank_model}
\end{minipage}

The simulation was performed in Simulink\textsuperscript{\textregistered}. For this example, 20 different settings were processed, whereof 16 served as training and four as validation sets.
The dynamics of the tank system is given in Eq. \eqref{eq:tank_dynamics}.
\begin{align}
\begin{split}
\label{eq:tank_dynamics}
	h(t+1) = h(t) + u(t-l_p) + Q_{\text{out}} \text{, with } \\
	u(t) =  \begin{cases}
	Q_{\text{in}}, & \text{if pump on}, \\
	0, & \text{if pump off.}
	\end{cases}
\end{split}
\end{align}
As a result, the pump works in rectangular pulses and the water level changes in a saw-tooth curve.

\subsection{Influence of polynomial degree}
\begin{figure}[htb]
\centering
\includegraphics[width=\linewidth]{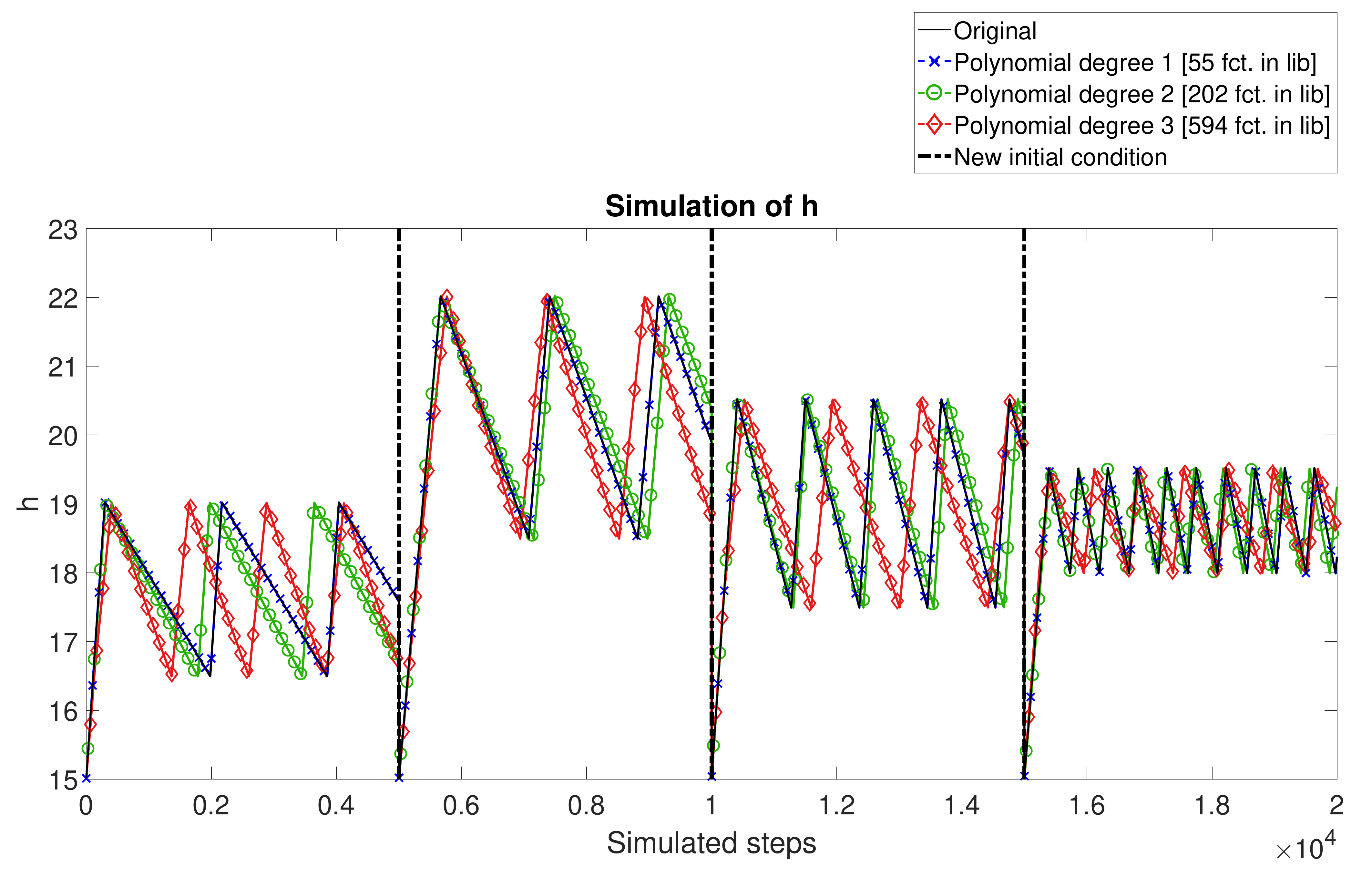}
	\captionof{figure}{Filling level for different degrees of polynomials}		
	\label{fig:exp2_sim_h} 
\end{figure}

The library prepared for SINDy consists of polynomials and hysterons. In a first experiment, we varied the degree of the polynomials. Fig. \ref{fig:exp2_sim_h} contains the simulated prediction of the filling level. Apparently, polynomials of low degree lead to the best results. For polynomials of degree one, the identified model is shown in Eq. \eqref{eq:poly1_model}. It consists of 4 out of 55 functions of the library, meaning a high degree of sparsity. Due to the hybrid form of hysteron $H_1$, it has the same structure as the correct model in Eq. \eqref{eq:tank_dynamics}. The models of higher degree manage to reproduce state transitions at the correct critical points, but in free running simulation, they induce an inaccuracy for the slope.\\
\begin{align}
		\tilde{h}_\text{poly1}(k+1) &= h(k)  + 1.56 \times 10^{-4}(4.98 Q_\text{in}  (1 - H_1) + Q_\text{out} ) \label{eq:poly1_model}, \nonumber \\
		H_{1}(k) &= \begin{cases}
		1, & \text{if } h > h_{\max}, \\
		0, & \text{if } h < h_{\min}, \\
		H_{1}(k-1), & \text{else}.
		\end{cases}	
\end{align}

\subsection{Effect of noisy data}
\begin{figure}[htb]
\centering
\includegraphics[width=\linewidth]{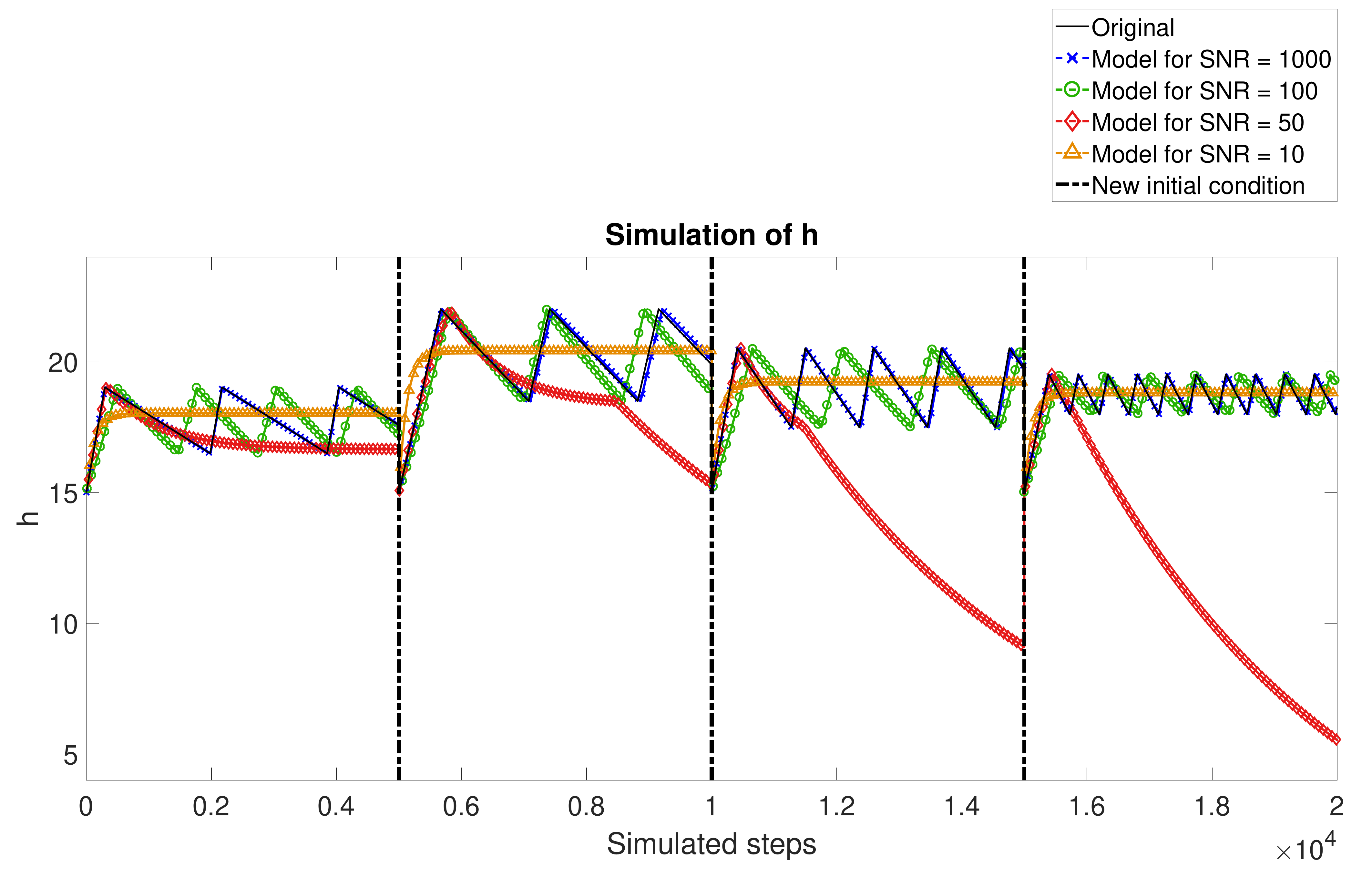}
\captionof{figure}{Filling level for different SNR}		
\label{fig:exp4_sim_h} 
\end{figure}

As a further objective, we analyzed the effect of noise. In this experiment we investigated  signal-to-noise ratios (SNR) 1000, 100, 50 and 10. The results are depicted in Fig. \ref{fig:exp4_sim_h}.
It can be seen that noise influences the identification quality very harshly. The approach works properly for an SNR of 1000. For an SNR of 100, the slope of the dynamics is not modeled accurately, resulting in a phase shift. The model for SNR 50 apparently identified two discrete states, but the state transitions as well as the dynamics within the states are faulty. For SNR 10, the measurements are too noisy to identify discrete states. As such, the model tries to satisfy all measurements by converging to the average.

\subsection{Influence of sample rate}
\begin{figure}[htb]
\centering
\includegraphics[width=\linewidth]{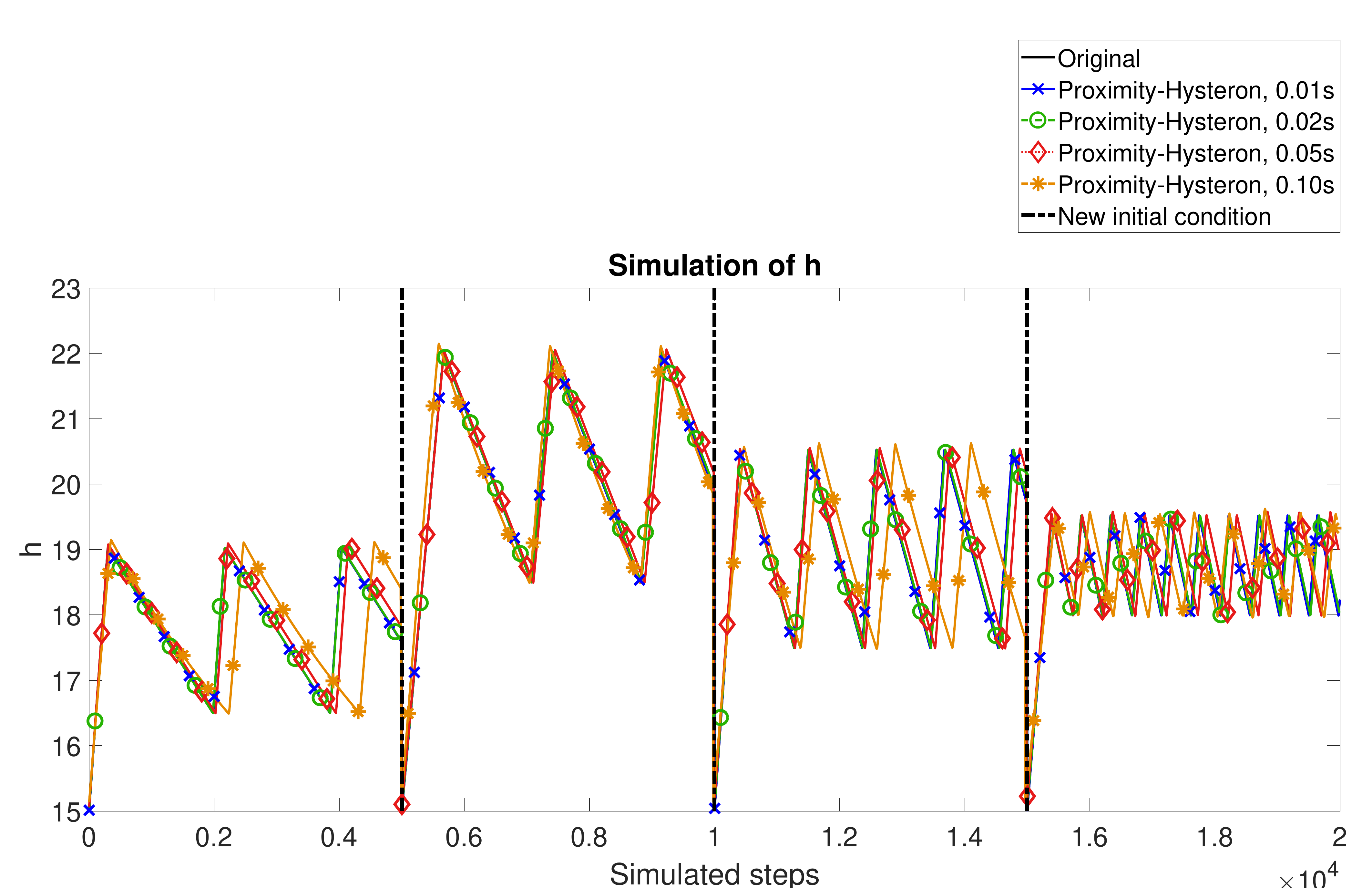}
	\captionof{figure}{Filling level for different sample rates with proximity hysteron}		
	\label{fig:exp3_proxi_sim_h} 
\end{figure}

Beside noise, an insufficient sampling rate is a typical problem for data-driven approaches. As explained in Sec. \ref{sec:proximity}, this method cannot localize transitions if the critical points are not included in the given data, which may happen for slow sampling rates. 
\begin{figure}[htb]
	\includegraphics[width=\linewidth]{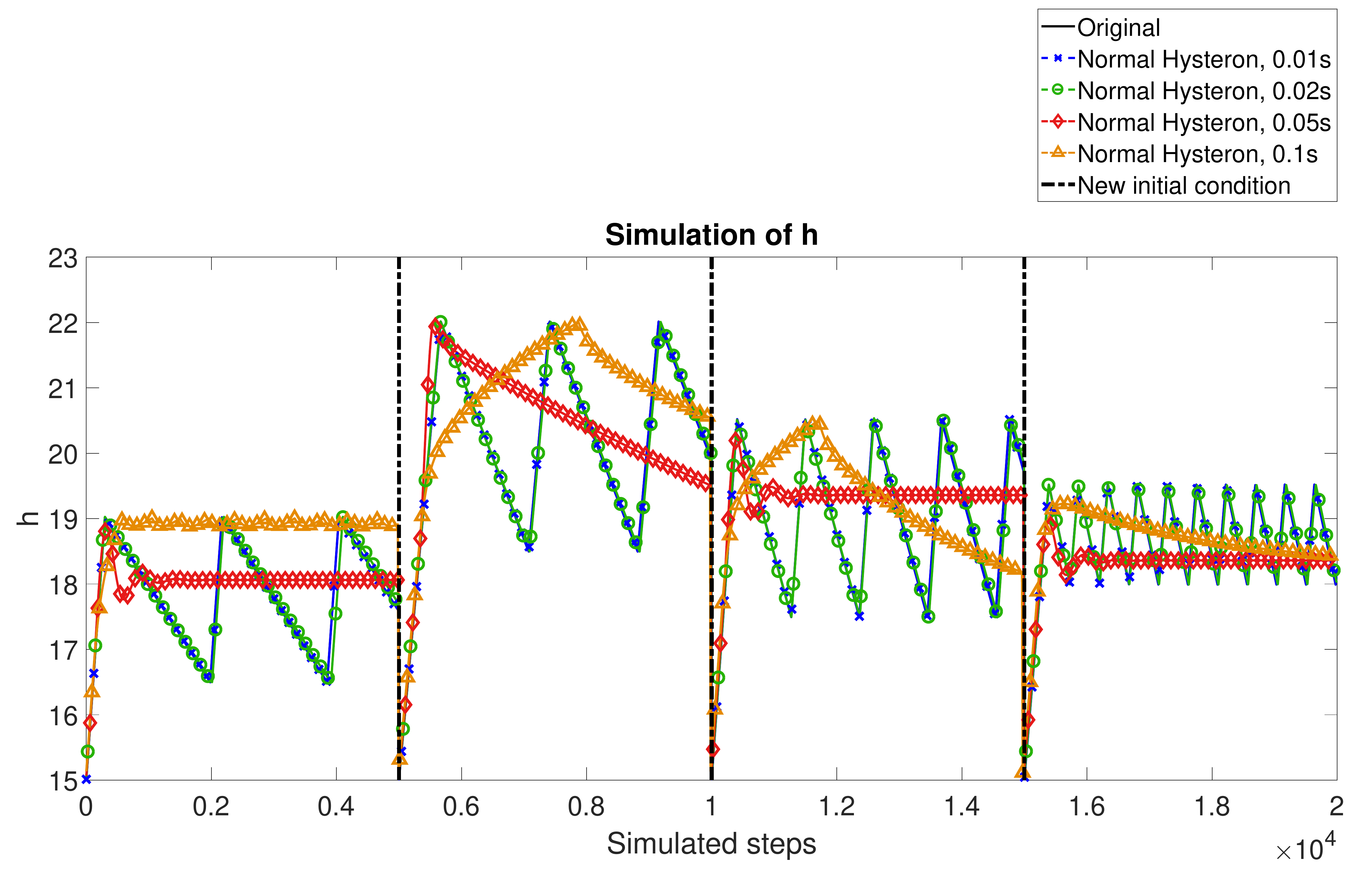}
	\captionof{figure}{Filling level for different sample rates with regular hysteron}		
	\label{fig:exp3_normal_sim_h} 
\end{figure}
Fig. \ref{fig:exp3_normal_sim_h} shows that state transitions are not detected using standard hysterons for slower sampling than $0.02$ s.
As a remedy, we presented the proximity hysteron in Sec. \ref{sec:proximity}. The effect can be seen in Fig. \ref{fig:exp3_proxi_sim_h}. With this extension, the algorithm achieves good results for all sampling rates. The best model is shown in Eq. \eqref{eq:snr_model} and was built from a library of 23 basis functions. It is equivalent to the model in Eq. \eqref{eq:poly1_model}, as $1 - H_1 = \bar{H}_1$. \\
\begin{align}
	\tilde{h}_\text{prox, 0.01}(k+1) &= h(k)  + 1.56 \times 10^{-4}(4.98 Q_\text{in} \ \bar{H}_1 + Q_\text{out} ) \label{eq:snr_model}, \nonumber \\
	H_{1}(k) &= \begin{cases}
		1, & \text{if } h > h_{\max}, \\
		0, & \text{if } h < h_{\min}, \\
		H_{1}(k-1), & \text{else}.
		\end{cases}	
\end{align}

\section{Discussion}
\label{sec:discussion}

The results show that SINDy with tailored basis functions in form of hysterons (SINDyHybrid) is a suitable method for the identification of hysteresis-controlled systems. The continuous system dynamics and the discrete states do not have to be modeled separately. With given signals and guesses for the thresholds, the hysteresis relation is found automatically. Adequate models were identified under varying sample rate and signal noise ratio. Designing a suited library remains the main challenge for the user. It was shown that systemic knowledge can be included in the function library and that specific adaptations, e.g. with regard to poor data quality, can be integrated. The library should not be larger than necessary, as this complicates the regression and hinders identification of the correct basis functions (which happens in the water basin problem when polynomials of a degree higher than 1 are used). 
The resulting models are sparse and thus easily interpretable. This enables testing for plausibility, which is required for many industrial applications. 
However, this method may show problems for systems with many discrete states. For each state, a hysteron will be identified, which is multiplied with functions describing the dynamics for that particular state. This results in a solution with a high number of overall terms, which is not according to the idea behind SINDy that the optimal solution is sparse. 
It must be noted that creating the hysteron requires knowledge about the actual threshold value of the state transition. For two-point controlled systems, those threshold values can usually be read from data. The $\epsilon$-range around these threshold values used by the proximity hysteron can compensate the discussed inaccuracies, and in this way even compensate aging effects. 
The presented experiments were performed using a MATLAB implementation of SINDy. In 2020, SINDy is also available for Python \cite{de_Silva2020}.

\section{Conclusion}
\label{sec:conclusion}
The presented work proposes a novel data-driven approach for identification of hybrid systems, SINDyHybrid. This method is exemplarily applied to a hysteresis-controlled system. The flexibility of the SINDy framework enables the integration of tailored basis functions, which can incorporate domain-specific knowledge. Based upon the relay-operator of the Preisach model, we developed the proximity hysteron, which allows for robust identification of state transitions and dynamics.

\bibliographystyle{IEEEtran}
\bibliography{IEEEabrv,literature}

\end{document}

%% file: tank_model.tex
% vorlage von hier : https://tex.stackexchange.com/questions/196680/how-can-i-draw-this-reservoir-diagram-with-tikz
%--------------------------

\colorlet{water}{cyan!25} % Define color for the water

% Dimensions of the tank
\def\tankwidth{6}
\def\tankheight{4}
\def\waterheight{2.2} % Water height 
%--------------------------
\tikzset{
	faucetIn/.pic={ % Define a 'pic' for the water inlet and outlet (PGF 3.0)
		\fill[water](-0.25,-0.25) rectangle (0.25,0.25);
		\draw[line width=1pt](-0.25,-0.25)--(0.25,-0.25) (-0.25,0.25)--(0.25,0.25);
	},
	myarrow/.tip={Stealth[scale=1.5]}, % Define a style for the tip of arrow
	surface water/.style= % style for border of water surface
	{decoration={random steps,segment length=1mm,amplitude=0.5mm}, decorate}
}
\tikzset{
	faucetOut/.pic={ % Define a 'pic' for the water inlet and outlet (PGF 3.0)
		\fill[water](-0.25,-0.125) rectangle (0.25,0.125);
		\draw[line width=1pt](-0.25,-0.125)--(0.25,-0.125) (-0.25,0.125)--(0.25,0.125);
	},
	myarrow/.tip={Stealth[scale=1.5]}, % Define a style for the tip of arrow
	surface water/.style= % style for border of water surface
	{decoration={random steps,segment length=1mm,amplitude=0.5mm}, decorate}
}

\begin{center}

\scalebox{0.9}{
\begin{tikzpicture}
% Water fill (I filled first so that way it is in the background)
\fill[water] decorate[surface water]{(\tankwidth,\waterheight) -- (0,\waterheight)}--(0,0) -- (\tankwidth,0) -- cycle;

% Tank
\draw[line width=1pt] (0,0) rectangle (\tankwidth,\tankheight);

\coordinate (entrance) at (0,\tankheight-0.7);
\coordinate (entranceLabel) at (-0.1,\tankheight-0.1);
\coordinate (exit) at (\tankwidth,0.7);
\coordinate (pumpLoc)at (-0.74,\tankheight-0.7);

\pic[xshift=-2.5mm+0.5pt] at (entrance) {faucetIn}; % water inlet (0.5pt is half of line width) 
\pic[xshift=2.5mm-0.5pt] at (exit) {faucetOut}; % outlet water

%\node[draw,circle] (pumpLoc) {Pump}; 
%\draw (pumpLoc.70) |- ++(1mm,5mm) -| (pumpLoc.east);

\draw [line width=.5pt, fill = white] (pumpLoc) circle (0.36cm);
\draw [line width=.5pt] (-0.74, \tankheight-0.34) -- (-0.38, \tankheight-0.7);
\draw [line width=.5pt] (-0.74, \tankheight-1.06) -- (-0.38, \tankheight-0.7);
\draw [|-|] (-0.5, \tankheight +0.25 ) -- (0, \tankheight +0.25) node[midway, above]{$l_\text{P}$};

% Entrance label (with `siunitx`)
\node[align=right,left=0cm] (inlet-unit) at (entranceLabel)  {Pump};
 %[align=...] in the last node is necessary for splitting in two lines with `\\`
%\draw[-myarrow](inlet-unit)--([xshift=-1mm]entrance);

% Exit label
\node[align=left,right=1cm] (outlet-unit) at (exit) {Drain};
\draw[-myarrow]([xshift=5mm]exit)--(outlet-unit);

\draw[|-|] ([xshift=-4mm]0,0) -- node[fill=white,inner xsep=0]{$h(t)$}([xshift=-4mm]0,\waterheight); 

% Fall water (i use `parabola` operation, it's more realistic, bacause it's a fall water) 
\fill[water] ([shift={(0.5pt,-2.5mm)}]entrance) parabola (0.3*\tankwidth,1pt) -- 
 (0.5*\tankwidth,1pt) parabola[bend at end] ([shift={(0.5pt,2.5mm)}]entrance);

\draw[dashed] (0,\waterheight+0.3) -- (\tankwidth,\waterheight+0.3) 
node [midway, above,fill=white] {$h_\text{max}$}; 
\draw[dashed] (0,\waterheight-0.3) -- (\tankwidth,\waterheight-0.3) 
node [midway, below,fill=water] {$h_\text{min}$};

%% Inner labels
%\path (0.5*\tankwidth,\tankheight)--(0.5*\tankwidth,0)
%    node[pos=0.2] {$x(t)$}
%    node[pos=0.5] {?\,L}
%    node[pos=0.8] {$x(0)=\SI{0}{\kilogram}$};
\end{tikzpicture}
}
\end{center}
\begin{flushleft}
	\captionof{figure}{Illustration of the modeled tank system}
	\label{fig:Wassertank}
\end{flushleft}